\documentclass[12pt]{article}
\usepackage{graphics}
\usepackage{graphicx}
\usepackage{amsfonts}
\usepackage{amsmath}
\usepackage[numbers,sort&compress]{natbib}
\begin{document}
\date{}

\title{Symbolic-computation study of integrable properties for the (2+1)-dimensional Gardner equation with the two-singular-manifold method}%
\author{ Hai-Qiang Zhang$^{1}$\hspace{0.3mm}, Juan Li$^{1}$, Tao Xu$^{1}$, Ya-Xing Zhang$^{1}$ and Bo
Tian$^{1,\,2}$\thanks{\textit{E-mail address:}
gaoyt@public.bta.net.cn (B.\ Tian), zhanghqbupt@\,yahoo.com.cn (H.\
Q.\ Zhang).}
\\
\\{\em 1. School of Science, P. O. Box 122, Beijing University of
Posts }\\{\em and Telecommunications, Beijing 100876, China}
\\{\em 2. Key Laboratory of Optical Communication and Lightwave Technologies,}\\
{\em Ministry of Education, Beijing University of Posts and}\\
{\em Telecommunications, Beijing 100876, China}} \maketitle

\begin{abstract}
The singular manifold method from the Painlev\'{e} analysis can be
used to investigate many important integrable properties for the
nonlinear partial differential equations. In this paper, the
two-singular-manifold method is applied to the (2+1)-dimensional
Gardner equation with two Painlev\'{e} expansion branches to
determine the Hirota bilinear form, B\"{a}cklund transformation, Lax
pairs and Darboux transformation.  Based on the obtained Lax pairs,
the binary Darboux transformation is constructed and the $N \times
N$ Grammian solution is also derived by performing the iterative
algorithm $N$ times with symbolic computation.
\end{abstract}
\newpage

\noindent\textbf{1. Introduction} \\\hspace*{\parindent}Arising from
the Painlev\'{e} analysis proposed by  Weiss, Tabor and
Carnevale~\cite{weiss83}, the singular manifold method (SMM) has
been successfully used to investigate the typical integrable
properties for many integrable nonlinear partial differential
equations (NPDEs), such as the Lax pair~\cite{weiss83b,MR91},
auto-B\"{a}cklund transformation~\cite{weiss83b,MR91,AM87},
nonclassical Lie symmetry~\cite{P92} and Hirota bilinear
formulation~\cite{WN1988}. In Refs.~\cite{AM87,FM89}, it is shown
the SMM has turned out to be capable of obtaining some special
classes of solutions for non-integrable NPDEs. However, due to the
existence of several Painlev\'{e} expansion branches for some given
NPDEs like the modified Korteweg-de Vries (mKdV)
equation~\cite{MR94}, Sine-Gordon (SG) equation~\cite{MR94} and
modified Kadomtsev-Petviashvili (KP) equation~\cite{PP97}, in this
situation the SMM is not feasible to exploit the integrable
properties of these equations. Therefore, Refs.~\cite{MR94,RM95,P93}
have generalized the SMM and developed the two-singular-manifold
method to uncover information about integrable character.

Different from the usual expansion, the two-singular-manifold method
involves two truncated Painlev\'{e} expansions at the constant level
term, which contains two different singular manifolds at a time.
This approach has been applied to the mKdV
equation~\cite{MR94,PG94}, SG equation~\cite{MR94}, classical
Boussinesq system~\cite{P93,RM95}, Mikhailov-Shabat
system~\cite{P93}, generalized dispersive long wave
equation~\cite{PP97}, modified KP equation~\cite{PP97}, and so on.
With this method, not only the auto-B\"{a}cklund transformation and
Lax pair can be obtained, but also the Darboux transformation can be
constructed in terms of the truncated Painlev\'{e} expansions in
both the NPDE and its Lax pair~\cite{PP97,JP98,P99}. In addition,
the relationship relating the singular manifolds and Hirota
$\tau$-function can be precisely
established~\cite{P93,RM95,PG94,PP97}.
\renewcommand{\theequation}{1.\arabic{equation}}
\setcounter{equation}{0}

Permeation of symbolic computation among various fields of science
and engineering remarkably helps the investigations on the nonlinear
partial differential equations (NPDEs)~\cite{GJ99,MJ04,Tian05,n2}.
Symbolic computation has increased the ability of a computer to deal
with a large amount of complicated and tedious algebraic
calculations.

In this paper, by virtue of the symbolic computation, we will
investigate the integrable properties for the (2+1)-dimensional
Gardner equation~\cite{BV84,B91}

\begin{equation}
u_t -u_{xxx}-6\,\beta\,u\,u_x+\frac{3}{2}\,\alpha^2\,u^2\,u_x-3\int
u_{yy}\,dx+3\,\alpha\,u_x\int u_y\,dx=0,\label{Gar1.2}
\end{equation}
where $\alpha$ and $\beta$ are two arbitrary constants. When
$u_y=0$, Eqn.~(\ref{Gar1.2}) reduces to the well-known
(1+1)-dimensional Gardner equation. For $\alpha=0$,
Eqn.~(\ref{Gar1.2}) is the KP equation, while it is the modified KP
equation with $\beta=0$. Therefore, the (2+1)-dimensional Gardner
equation could be regarded as a combined KP and modified KP
equation. Eqn.~(\ref{Gar1.2}) is completely integrable in the sense
that it has been solved by the inverse spectral transform
method~\cite{B91}. Refs.~\cite{B91,XC01,I99,YZ05,GH07} have
presented its wide classes of analytical solutions including the
rational solution, quasi-periodic solution , soliton solution and
non-decaying real solutions.

In the following sections, with the help of symbolic computation, we
will apply the two-singular-manifold method to the (2+1)-dimensional
Gardner equation to  determine the Hirota bilinear form,
B\"{a}cklund transformation and Lax pairs. Based on the obtained Lax
pairs, we will construct the binary Darboux transformation and
perform symbolic computation on the iterative algorithm to generate
the Grammian solutions.
\\
\\
\noindent\textbf{2. Hirota bilinear form}
\\\hspace*{\parindent}To begin with, we rewrite Eqn.~(\ref{Gar1.2}) as
the following system
\renewcommand{\theequation}{2.\arabic{equation}}
\setcounter{equation}{0}
\begin{subequations}
\begin{align}
&u_t
-u_{xxx}-6\,\beta\,u\,u_x+\frac{3}{2}\,\alpha^2\,u^2\,u_x-3\,v_y+3\,\alpha\,u_x\,v=0,
\\
&v_x=u_y.
\end{align}\label{Gar2.1}
\end{subequations}
Then, we expand the solutions of System~(\ref{Gar2.1}) in a
generalized Laurent series
\begin{eqnarray}
u=\sum_{j=0}^{\infty}u_{j}\,\chi^{-a+j},\ \ \ \ \ \ \
v=\sum_{j=0}^{\infty}v_{j}\,\chi^{-b+j},\label{Gar2.2}
\end{eqnarray}
where $\chi=\chi(x,y,t)$, and $u_j=u_j(x,y,t)$, $v_j=v_j(x,y,t)$ are
analytical functions in the neighborhood of a non-characteristic
movable singularity manifold $\chi(x,y,t)=0$, while $a$ and $b$ are
two integers to be determined. By the analysis of the leading terms,
we obtain
\begin{eqnarray}
a = 1,\ \ b = 1,\ \ u_0 =2\,\epsilon \frac{\chi_x}{\alpha},\ \ v_0
=2\,\epsilon \frac{\chi_y}{\alpha},\label{Gar2.3}
\end{eqnarray}
where $\epsilon=\pm1$. It is easy to see that  $u_0$ and $v_0$ can
take two values so that System~(\ref{Gar2.1}) has two different
Painlev\'{e} expansion branches. By using two different singular
manifolds $\phi$ and $\varphi$~\cite{PP97,P99,JP98}, we take the
truncated Painlev\'{e} expansion at the constant level term
\begin{subequations}
\begin{align}
&u'=u+\frac{2}{\alpha}\left(\frac{\phi_x}{\phi}-\frac{\varphi_x}{\varphi}\right),
\\&v'=v+\frac{2}{\alpha}\left(\frac{\phi_y}{\phi}-\frac{\varphi_y}{\varphi}\right),
\end{align}\label{Gar2.4}
\end{subequations}\hspace{-2mm}where the singular manifold $\phi$ corresponds
to $\epsilon = 1$ and $\varphi$ to $\epsilon =-1$, which can also be
regarded as an auto-B\"{a}cklund transformation between two
different solutions $(u', v')$ and $(u, v)$ for
System~(\ref{Gar2.1}), when singular manifolds $\phi$ and $\varphi$
satisfy the truncation conditions.

Motivated by Expressions (\ref{Gar2.4}), we introduce the dependent
variable transformations
\begin{subequations}
\begin{align}
&u=\frac{2}{\alpha}\left(\frac{g_x}{g}-\frac{f_x}{f}\right)=\frac{2}{\alpha}\left({\rm{log}}\frac{g}{f}\right)_x,\label{Gar2.4b}
\\&v=\frac{2}{\alpha}\left(\frac{g_y}{g}-\frac{f_y}{f}\right)=\frac{2}{\alpha}\left({\rm{log}}\frac{g}{f}\right)_y,\label{Gar2.5b}
\end{align}\label{Gar2.5}
\end{subequations}\vspace{1cm}\hspace{-1.5mm}to transform System~(\ref{Gar2.1}) into the Hirota bilinear form.
Substituting Expressions~(\ref{Gar2.5}) back into
System~(\ref{Gar2.1}), we obtain
\begin{eqnarray}
&&\hspace{-6mm}\left(\frac{D_t g \cdot f}{g f}-\frac{D_x^3 g \cdot
f}{g f}\right)_x+3\left(\frac{D_x^2 g \cdot f}{g
f}\right)_x\,\frac{D_x g \cdot f}{g f}+3\,\frac{D_x^2 g \cdot f}{g
f}\,\left(\frac{D_x g \cdot f}{g f}\right)_x-2\left(\frac{D_x g
\cdot f}{g f}\right)_x^3\nonumber
\\&&-\frac{6\beta}{\alpha}\left(\frac{D_x g \cdot f}{g
f}\right)_x^2+2\left(\frac{D_x g \cdot f}{g
f}\right)_x^3-3\left(\frac{D_y g \cdot f}{g f}\right)_y+6\frac{D_y g
\cdot f}{g f}\,\left(\frac{D_x g \cdot f}{g f}\right)_x =
0,\label{Gar2.6}
\end{eqnarray}where $D$ is the well-known Hirota bilinear
operator~\cite{Hirota}
\begin{eqnarray}
D_x^mD_y^nD_t^l\,\,g \cdot f = (\partial_x
-\partial_{x'})^m(\partial_y -\partial_{y'})^n(\partial_t
-\partial_{t'})^lg(x,y,t)f(x',y',t')|_{x' = x,\, y' = y,\, t'
=t}.\label{Gar2.7}
\end{eqnarray}
With symbolic computation,  Eqn.~(\ref{Gar2.6}) can be split into
\begin{eqnarray}
&&\left(D_y + D_x^2-2\,\beta / \alpha\, D_y\right)\,g \cdot f = 0,\label{Gar2.8}\\
&&\left(D_t-D_x^3+3\,D_xD_y-6\,\beta / \alpha\, D_y\right)\,g \cdot
f = 0,\label{Gar2.9}
\end{eqnarray}
which are the Hirota bilinear form of Eqn.~(\ref{Gar1.2}). By the
perturbation technique, one can assume the functions $f$ and $g$ in
powers of a small parameter $\varepsilon$~\cite{Hirota} to obtain
the multi-soliton solutions of Eqn.~(\ref{Gar1.2}) from
Eqns.~(\ref{Gar2.8}) and (\ref{Gar2.9}).

It is noted that the key step for the Hirota method is to seek for
the suitable dependent variable transformation for a given NPDE to
be transformed into the Hirota bilinear form. If we do not know how
to do this, then there is little prospect of being able to use the
Hirota method. However, the truncated expansion in Painlev\'{e}
analysis can provide us with a useful clue in finding such desired
transformations. In fact, the Hirota bilinear forms for a large
class of NPDEs can be obtained in terms of the Painlev\'{e}
truncated expansion~\cite{TY05,Lak87,Lak95}.
\\
\\
\noindent\textbf{3. Bilinear B\"{a}cklund transformation}
\\\hspace*{\parindent}For an integrable NPDE, the existence of a B\"{a}cklund transformation
seems to be widely accepted~\cite{Zabook,R82}. In this section, from
Eqns.~(\ref{Gar2.8}) and (\ref{Gar2.9}), we will derive a bilinear
B\"{a}cklund transformation between two different solutions
$u=\frac{2}{\alpha}\left({\rm{log}}\,g/f\right)_x$ and
$u'=\frac{2}{\alpha}\left({\rm{log}}\,g'/f'\right)_x$ for
Eqn.~(\ref{Gar1.2}),
\renewcommand{\theequation}{3.\arabic{equation}}
\setcounter{equation}{0}by considering the following two equations,
\begin{eqnarray}
&&P1=\left[\left(D_y + D_x^2-2\,\beta / \alpha\, D_y\right)g \cdot
f\right]g'\,f'-g\,f\left[\left(D_y + D_x^2-2\,\beta / \alpha\,
D_y\right)g' \cdot f'\right],\label{GarB.1}
\\ \nonumber&&P2=\left[\left(D_t-D_x^3+3\,D_xD_y-6\,\beta / \alpha\, D_y\right)\,g \cdot
f\right]g'\,f'
\\&&\ \ \ \ \ \ \ \
-g\,f\left[\left(D_t-D_x^3+3\,D_xD_y-6\,\beta / \alpha\,
D_y\right)\,g' \cdot f'\right]\label{GarB.2}.
\end{eqnarray}
With the aid of the Hirota bilinear operator identities (see
Appendix (\ref{GarA1})$-$(\ref{GarA5})), symbolic computation on
Eqns.~(\ref{GarB.1}) and (\ref{GarB.2}) yields
\begin{eqnarray}
&& P1=\left[\left(D_y + D_x^2-2\,\beta / \alpha\, D_y\right)g \cdot
g'\right]f\,f'-g\,g'\left[\left(D_y + D_x^2-2\,\beta / \alpha\,
D_y\right)f \cdot f'\right]\nonumber
\\&& \hspace{1.1cm}-2\,D_x\,(g\,f')\cdot (D_x\,f\cdot
g')\label{GarB.3},
\\&&P2=3\,D_x(D_x\,g\cdot f')\cdot (D_x\,f\cdot g')-3\,D_x(g\,f')\cdot (D_y\,f\cdot
g')-3\,D_y\,(g\,f')\cdot(D_x\,f\cdot g') \nonumber
\\&&\ \ \ \ \ \ \ \ +\left[\left(D_t-D_x^3+3\,D_xD_y-6\,\beta / \alpha\, D_y\right)\,g
\cdot g'\right]f\,f'\nonumber
\\&&\ \ \ \ \ \ \ \ -\left[\left(D_t-D_x^3+3\,D_xD_y-6\,\beta / \alpha\, D_y\right)\,f
\cdot f'\right]g\,g' \label{GarB.4}.
\end{eqnarray}Thus, Eqns.~(\ref{GarB.3}) and (\ref{GarB.4}) can be further decoupled into
the following equations
\begin{subequations}
\begin{align}
&D_xf\cdot g'=\eta(t)\,g\,f',\label{GarB.5a}
\\&\eta(t)\,D_x\,g\cdot f' + D_y\,f\cdot g' +\gamma(t)\,g\,f'=0,\label{GarB.5b}
\\&\left[D_y + D_x^2-2\,\beta / \alpha\, D_y+\xi(t)\right]g\cdot g'=0,\label{GarB.5c}
\\&\left[D_y + D_x^2-2\,\beta / \alpha\, D_y+\xi(t)\right]f\cdot
f'=0,\label{GarB.5d}
\\&\left[D_t-D_x^3+3\,D_xD_y-6\,\beta / \alpha\, D_y+\zeta(t)\right]g\cdot g'=0,\label{GarB.5e}
\\&\left[D_t-D_x^3+3\,D_xD_y-6\,\beta / \alpha\, D_y+\zeta(t)\right]f\cdot\hspace{-0.5mm}f'=0,\label{GarB.5f}
\end{align}\label{GarB.5}
\end{subequations}\hspace{-1.5mm}where $\eta(t)$, $\gamma(t)$,
$\xi(t)$ and $\zeta(t)$ are all arbitrary  differentiable functions
of $t$. Eqns.~(\ref{GarB.5}) constitute the bilinear B\"{a}cklund
transformation for Eqn.~(\ref{Gar1.2}), from which more complicated
solutions can be progressively constructed beginning with a seed
solution. Additionally, it can also be of use for the investigation
on  other integrable
properties~\cite{Hirota,Na81,R82,Zhang06,Zabook,Tian07,me1}, like
the nonlinear superposition formula, Lax pair, conservation laws,
etc.
\\
\\
\noindent\textbf{4. Lax pairs with symbolic computation}
\\\hspace*{\parindent}In this section, by the two-singular-manifold method, the
Lax pairs of the (2+1)-dimensional Gardner equation will be derived.
With symbolic computation, we insert Expressions~(\ref{Gar2.4}) into
System~(\ref{Gar2.1}), and get
\renewcommand{\theequation}{4.\arabic{equation}}
\setcounter{equation}{0}
\begin{eqnarray}
&&\frac{\phi_x}{\phi}\,\frac{\varphi_x}{\varphi}=
A\,\frac{\phi_x}{\phi}+B\,\frac{\varphi_x}{\varphi},\label{Gar3.1}
\\&&
12\,\beta\,u-3\,\alpha^2\,u^2+2\,v_1^2-6\,\alpha\,v-2\,w_1+6\,\tau_1^2+6\,\alpha\,u_x+8\,v_{1x}=0,\label{Gar3.2}
\\&&
12\,\beta\,u-3\,\alpha^2\,u^2+2\,v_2^2-6\,\alpha\,v-2\,w_2+6\,\tau_2^2-6\,\alpha\,u_x+8\,v_{2x}=0,\label{Gar3.3}
\\&&3\,\tau_{1y}-v_1\,v_{1x}-w_{1x}+3\,\tau_1\,\tau_{1x}+v_{1xx}=0,\label{Gar3.4}
\\&&3\,\tau_{2y}-v_2\,v_{2x}-w_{2x}+3\,\tau_2\,\tau_{2x}+v_{2xx}=0,\label{Gar3.5}
\end{eqnarray}
with
\begin{eqnarray} \hspace{-200mm}&& A=\frac{1}{2\,\alpha}\left( \alpha\,
v_1-2\,\beta+\alpha^2\,u+\alpha\,\tau_1\right),\label{Gar3.6}
\\&& B=\frac{1}{2\,\alpha}\left( \alpha\,
v_2+2\,\beta-\alpha^2\,u-\alpha\,\tau_2\right),\label{Gar3.7}
\end{eqnarray}
where $v_i$, $w_i$ and $\tau_i$ $(\,i=1,2\,)$ are defined as
\begin{eqnarray}
&&v_1=\frac{\phi_{xx}}{\phi_x},\ \ \ \ \ \ \ \ \
v_2=\frac{\varphi_{xx}}{\varphi_x},\label{Gar3.8}
\\&&w_1=\frac{\phi_{t}}{\phi_x},\ \ \ \ \ \ \ \ \
w_2=\frac{\varphi_{t}}{\varphi_x},\label{Gar3.9}
\\&&\tau_1\,=\frac{\phi_{y}}{\phi_x},\ \ \ \ \ \ \ \ \
\,\tau_2\,=\frac{\varphi_{y}}{\varphi_x}.\label{Gar3.10}
\end{eqnarray}
We use the derivatives of Expression~(\ref{Gar3.1}) with respect to
$x$, $y$ and $t$, so as to obtain
\begin{eqnarray}
&&A_x=A\left(v_2-A-B\right),\label{Gar3.11}
\\&&A_y=\left(A\,\tau_{2}\right)_x+AB(\tau_2-\tau_1),\label{Gar3.12}
\\&&A_t=\left(A\,w_{2}\right)_x+AB(w_2-w_1),\label{Gar3.13}
\\&&B_x=B\left(v_1-A-B\right),\label{Gar3.14}
\\&& B_y=\left(B\,\tau_{1}\right)_x-AB(\tau_2-\tau_1),\label{Gar3.15}
\\&&B_t=\left(B\,w_{1}\right)_x-AB(w_2-w_1).\label{Gar3.16}
\end{eqnarray}
By virtue of Eqns.~(\ref{Gar3.6})$-$(\ref{Gar3.16}), it is easy to
check that the following two relationships are satisfied
\begin{eqnarray}
&&\left(AB\right)_x=AB(\tau_2-\tau_1),\label{Gar3.17}
\\&&\left[AB(2\,A-2\,B+v_1-v_2-3\,\tau_1-3\,\tau_2)\right]_x=
AB(w_1-w_2).\label{Gar3.18}
\end{eqnarray}
Therefore, Eqns.~(\ref{Gar3.11})$-$(\ref{Gar3.16}) can be simplified
as
\begin{eqnarray}
&&A_x=A\left(v_2-A-B\right),\label{Gar3.19}
\\&&A_y=\left[A(\tau_{2}+B)\right]_x,\label{Gar3.20}
\\&&A_t=\left[A\,w_{2}-AB(2\,A-2\,B+v_1-v_2-3\,\tau_1-3\,\tau_2)\right]_x,\label{Gar3.21}
\\&&B_x=B\left(v_1-A-B\right),\label{Gar3.22}
\\&&B_y=\left[B(\tau_{1}-A)\right]_x,\label{Gar3.23}
\\&&B_t=\left[B\,w_{1}+AB(2\,A-2\,B+v_1-v_2-3\,\tau_1-3\,\tau_2)\right]_x.\label{Gar3.24}
\end{eqnarray}
Through introducing the changes as
\begin{eqnarray}
A=\frac{\psi_{x}^+}{\psi^+},\ \ \ \ \ \ \ \ \
B=\frac{\psi_{x}^-}{\psi^-},\label{Gar3.25}
\end{eqnarray}
Eqns.~(\ref{Gar3.19})$-$(\ref{Gar3.24}) can be linearized into
\begin{eqnarray}
&&\psi_{xx}^-=\psi_{x}^-(v_1-A),\label{Gar3.26}
\\&&\psi_{y}^-\,\,=\psi_{x}^-(\tau_1-A),\label{Gar3.27}
\\&&\psi_{t}^-\,\,=\psi_{x}^-[w_1+A(2\,A-2\,B+v_1-v_2-3\,\tau_1-3\,\tau_2)],\label{Gar3.28}
\\&&\psi_{xx}^+=\psi_{x}^+(v_2-B),\label{Gar3.29}
\\&&\psi_{y}^+\,\,=\psi_{x}^+(\tau_2+B),\label{Gar3.30}
\\&&\psi_{t}^+\,\,=\psi_{x}^+[w_2-B(2\,A-2\,B+v_1-v_2-3\,\tau_1-3\,\tau_2)].\label{Gar3.31}
\end{eqnarray}
Symbolic computation on Eqns.~(\ref{Gar3.26})$-$(\ref{Gar3.31}) with
the substitution of Eqns.~(\ref{Gar3.2}) and (\ref{Gar3.3}) gives
rise to
\begin{subequations}
\begin{align}
&\hspace{-2.2mm}\psi_{y}^-=-\psi_{xx}^--\left(\alpha\,u-\frac{2\,\beta}{\alpha}\right)\psi_{x}^-,
\\ &\hspace{-2.2mm}\psi_{t}^-=4\,\psi_{xxx}^-+\left(6\,\alpha\,u-\frac{12\,\beta}{\alpha}\right)\psi_{xx}^-+
\left(\frac{12\,\beta^2}{\alpha^2}-6\,\beta\,u+\frac{3}{2}\,\alpha^2\,u^2-3\,\alpha\,v+3\,\alpha\,u_x\right)\psi_{x}^-,
\end{align}\label{Gar3.32}
\end{subequations}
and
\begin{subequations}
\begin{align}
&\hspace{-2.5mm}\psi_{y}^+=\psi_{xx}^+-\left(\alpha\,u-\frac{2\,\beta}{\alpha}\right)\psi_{x}^+,
\\
&\hspace{-2.5mm}\psi_{t}^+=4\,\psi_{xxx}^+-\left(6\,\alpha\,u-\frac{12\,\beta}{\alpha}\right)\psi_{xx}^++
\left(\frac{12\,\beta^2}{\alpha^2}-6\,\beta\,u+\frac{3}{2}\,\alpha^2\,u^2-3\,\alpha\,v-3\,\alpha\,u_x\right)\psi_{x}^+.
\end{align}\label{Gar3.33}
\end{subequations}

By direct calculation, it is found that Eqn.~(\ref{Gar1.2}) can be
derived from the compatibility conditions
$\psi_{yt}^\pm=\psi_{ty}^\pm$.  Thus, Eqns.~(\ref{Gar3.32}) and
(\ref{Gar3.33}) are two different types of Lax pairs of the
(2+1)-dimensional Gardner equation. It is noted that through the
following gauge transformations
\begin{eqnarray}
&&\psi^-={\rm{exp}}\left\{\frac{\beta}{\alpha}\,x+\frac{\beta^2}{\alpha^2}\,y+\frac{\beta^3}{\alpha^3}\,t\right\}\,\Gamma^-,\nonumber
\\&&\psi^+={\rm{exp}}\left\{-\frac{\beta}{\alpha}\,x-\frac{\beta^2}{\alpha^2}\,y-\frac{\beta^3}{\alpha^3}\,t\right\}\,\Gamma^+,\nonumber
\end{eqnarray}
Lax pairs~(\ref{Gar3.32}) and (\ref{Gar3.33}) can be respectively
transformed into
\begin{subequations}
\begin{align}
&
\Gamma_{y}^-=-\Gamma_{xx}^--\alpha\,u\,\Gamma_{x}^--\beta\,u\,\Gamma^-,
\\& \Gamma_{t}^-=4\,\Gamma_{xxx}^-+6\,\alpha\,u\,\Gamma_{xx}^-+
\left(6\,\beta\,u+\frac{3}{2}\,\alpha^2\,u^2-3\,\alpha\,v+3\,\alpha\,u_x\right)\Gamma_{x}^-\nonumber
\\ & \ \ \ \ \ \ \ \
+\left(3\,\beta\,u_x+\frac{3}{2}\,\alpha\,\beta\,u^2-3\,\beta\,v\right)\Gamma^-,
\end{align}\label{Gar3.34}
\end{subequations}
and
\begin{subequations}
\begin{align}
&\Gamma_{y}^+=\Gamma_{xx}^+-\alpha\,u\,\Gamma_{x}^++\beta\,u\,\Gamma^+,
\\ \nonumber
&\Gamma_{t}^+=4\,\Gamma_{xxx}^+-6\,\alpha\,u\,\Gamma_{xx}^++
\left(6\,\beta\,u+\frac{3}{2}\,\alpha^2\,u^2-3\,\alpha\,v-3\,\alpha\,u_x\right)\Gamma_{x}^+
\\ & \ \ \ \ \ \ \ \ +\left(3\,\beta\,u_x-\frac{3}{2}\,\alpha\,\beta\,u^2+3\,\beta\,v\right)\Gamma^+.
\end{align}\label{Gar3.35}
\end{subequations}\hspace{-2mm}The compatibility conditions $\Gamma_{yt}^\pm=\Gamma_{ty}^\pm$ can
also give rise to Eqn.~(\ref{Gar1.2}). Note that this form of Lax
pair (\ref{Gar3.34}) has been presented in Ref.~\cite{BV84}, and
other three Lax pairs are given here for the first time.

To this stage, with the two-singular-manifold method, we have
obtained the Lax pairs of the (2+1)-dimensional Gardner equation,
i.e., Systems (\ref{Gar3.32})$-$(\ref{Gar3.35}). In the next
section, we will construct the relationship between the singular
manifolds $\phi$, $\varphi$ and eigenfunctions $\psi^-$, $\psi^+$.
\\
\\
\noindent\textbf{5. Relationship between the singular manifolds and
eigenfunctions}
\\\hspace*{\parindent}Using Eqns.~(\ref{Gar3.8})$-$(\ref{Gar3.10}) and
(\ref{Gar3.25}), we rewrite  Eqns.~(\ref{Gar3.26})$-$(\ref{Gar3.31})
as
\renewcommand{\theequation}{5.\arabic{equation}}
\setcounter{equation}{0}
\begin{eqnarray}
&&\frac{\psi_{xx}^-}{\psi_{x}^-}=\frac{\phi_{xx}}{\phi_x}-\frac{\psi_{x}^+}{\psi^+},\label{Gar4.1}
\\&&\frac{\psi_{xx}^+}{\psi_{x}^+}=\frac{\varphi_{xx}}{\varphi_x}-\frac{\psi_{x}^-}{\psi^-},\label{Gar4.2}
\\&&\frac{\psi_{y}^-}{\psi_{x}^-}=\frac{\phi_{y}}{\phi_x}-\frac{\psi_{x}^+}{\psi^+},\label{Gar4.3}
\\&&\frac{\psi_{y}^+}{\psi_{x}^+}=\frac{\varphi_{y}}{\varphi_x}+\frac{\psi_{x}^-}{\psi^-},\label{Gar4.4}
\\&&\frac{\psi_{t}^-}{\psi_{x}^-}=\frac{\phi_{t}}{\phi_x}+\frac{\psi_{x}^+}{\psi^+}
\left(2\frac{\psi_{x}^+}{\psi^+}-2\frac{\psi_{x}^-}{\psi^-}+\frac{\phi_{xx}}{\phi_x}-\frac{\varphi_{xx}}{\varphi_x}-3\frac{\phi_{y}}{\phi_x}-3\frac{\varphi_{y}}{\varphi_x}\right),\label{Gar4.5}
\\&&\frac{\psi_{t}^+}{\psi_{x}^+}=\frac{\varphi_{t}}{\varphi_x}-\frac{\psi_{x}^-}{\psi^-}
\left(2\frac{\psi_{x}^+}{\psi^+}-2\frac{\psi_{x}^-}{\psi^-}+\frac{\phi_{xx}}{\phi_x}-\frac{\varphi_{xx}}{\varphi_x}-3\frac{\phi_{y}}{\phi_x}-3\frac{\varphi_{y}}{\varphi_x}\right).\label{Gar4.6}
\end{eqnarray}
Integrating Eqns.~(\ref{Gar4.1}) and (\ref{Gar4.2}) with respect to
$x$ yields
\begin{eqnarray}
&&\phi_x=\psi_{x}^-\,\psi^+,\label{Gar4.7}
\\&&\varphi_x=\psi_{x}^+\,\psi^-.\label{Gar4.8}
\end{eqnarray}
Then, by substituting Eqns.~(\ref{Gar4.7}) and (\ref{Gar4.8}) into
Eqns.~(\ref{Gar4.3})$-$(\ref{Gar4.6}), we obtain
\begin{eqnarray}
&&\phi_y=\psi_{y}^-\,\psi^++\psi_{x}^-\,\psi_{x}^+,\label{Gar4.9}
\\&&\varphi_y=\psi_{y}^+\,\psi^--\psi_{x}^-\,\psi_{x}^+,\label{Gar4.10}
\\&&\phi_t=\psi^+\,\psi_{t}^--2\,\psi_{xx}^-\,\psi^+_x+2\,\psi_{x}^+\,\psi_{y}^-+4\,\psi_{x}^-\,\psi_{y}^+,\label{Gar4.11}
\\&&\varphi_t=\psi^-\,\psi_{t}^++2\,\psi_{xx}^-\,\psi^+_x-2\,\psi_{x}^+\,\psi_{y}^--4\,\psi_{x}^-\,\psi_{y}^+.\label{Gar4.12}
\end{eqnarray}
From Eqns.~(\ref{Gar4.7})$-$(\ref{Gar4.12}), it is shown that the
singular manifolds $\phi$ and $\varphi$ are determined by the
eigenfunctions $\psi^-$ and $\psi^+$. So, by defining the singular
manifolds $\phi$ and $\varphi$ in the abbreviated form
as~\cite{PP97}
\begin{eqnarray}
&&\phi=\Delta(\psi^-,\psi^+),\label{Gar4.13}
\\&&\varphi=\Omega\,(\psi^-,\psi^+),\label{Gar4.14}
\end{eqnarray}\vspace{1cm}
Eqns.~(\ref{Gar4.7})$-$(\ref{Gar4.12}) can be written as
\begin{eqnarray}
&&\left[\Delta(\psi^-,\psi^+)\right]_x=\psi_{x}^-\,\psi^+,\label{Gar4.15}
\\&&\left[\Omega\,(\psi^-,\psi^+)\right]_x=\psi^-\,\psi_{x}^+,\label{Gar4.16}
\\&&\left[\Delta(\psi^-,\psi^+)\right]_y=\psi_{y}^-\,\psi^++\psi_{x}^-\,\psi_{x}^+,\label{Gar4.17}
\\&&\left[\Omega\,(\psi^-,\psi^+)\right]_y=\psi^-\,\psi_{y}^+-\psi_{x}^-\,\psi_{x}^+,\label{Gar4.18}
\\&&\left[\Delta(\psi^-,\psi^+)\right]_t=\psi^+\,\psi_{t}^--2\,\psi_{xx}^-\,\psi^+_x+2\,\psi_{x}^+\,\psi_{y}^-+4\,\psi_{x}^-\,\psi_{y}^+,\label{Gar4.19}
\\&&\left[\Omega\,(\psi^-,\psi^+)\right]_t=\psi^-\,\psi_{t}^++2\,\psi_{xx}^-\,\psi^+_x-2\,\psi_{x}^+\,\psi_{y}^--4\,\psi_{x}^-\,\psi_{y}^+.\label{Gar4.20}
\end{eqnarray}
Here, we can see that Eqns.~(\ref{Gar4.15})$-$(\ref{Gar4.20})
establish the relationship between two singular manifolds $\phi$,
$\varphi$ and eigenfunctions $\psi^-$, $\psi^+$, and possess the
following relation
\begin{eqnarray}
\Delta(\psi^-,\psi^+)+\Omega(\psi^-,\psi^+)=\psi^-\,\psi^+.\label{Gar4.21}
\end{eqnarray}
\\
\noindent\textbf{6. Binary Darboux transformation}
\\\hspace*{\parindent}The Darboux transformation method is a powerful tool to get the analytical solutions
for the integrable NPDEs~\cite{VM91,Gu05}. The most obvious
advantage of this method lies in its iterative algorithm, which is
purely algebraic and can be easily achieved on the symbolic
computation system. By virtue of the SMM and Darboux transformation,
starting from the seed solution and solving the corresponding linear
equation or system, one can obtain wide classes of exact analytical
solutions for a NPDE, such as the soliton solutions, periodic
solutions and rational solutions~\cite{VM91,Gu05,LZ03,Zh98,PS01,
AC00}.

Based on the Lax pairs (\ref{Gar3.32}) and (\ref{Gar3.33}) obtained
in Section 3, we can construct the binary Darboux transformation of
Eqn.~(\ref{Gar1.2}). Although Eqns.~(\ref{Gar2.4}) and
(\ref{Gar2.5}) are considered as an auto-B\"{a}cklund
transformation, it is actually not convenient to generate more and
more complicated solutions in a recursive manner, because it only
involves the transformation for potentials. In comparison, the
Darboux transformation not only has the potential transformation,
but also establishes the relationship between the new and old
eigenfunctions. Next, we turn our attention to the transformation of
eigenfunctions.

\renewcommand{\theequation}{6.\arabic{equation}}
\setcounter{equation}{0}

Let us assume that
\begin{eqnarray}
&&\psi^{'-}=\psi_1^-+\frac{f_1}{\phi}+\frac{f_2}{\varphi},\label{Gar5.1}
\\&&\psi^{'+}=\psi_1^++\frac{g_1}{\phi}+\frac{g_2}{\varphi},\label{Gar5.2}
\end{eqnarray}
where $\psi_1^-$ and $\psi_1^+$ respectively correspond to  the
solutions of Lax pairs~(\ref{Gar3.32}) and (\ref{Gar3.33}), while
$f_i$ and $g_i$ $(\,i=1,2\,)$ are two differentiable functions to be
determined, the new eigenfunctions $\psi^{'-}$ and $\psi^{'+}$ also
satisfy  Lax pairs~(\ref{Gar3.32}) and (\ref{Gar3.33}) except that
$(u,v)$ is replaced by $(u',v')$, namely,
\begin{subequations}
\begin{align}
&\psi_{y}^{'-}=-\psi_{xx}^{'-}-\left(\alpha\,u'-\frac{2\,\beta}{\alpha}\right)\psi_{x}^{'-},
\\ \nonumber
&\psi_{t}^{'-}=\left(\frac{12\,\beta^2}{\alpha^2}-6\,\beta\,u'+\frac{3}{2}\,\alpha^2\,u'^2-3\,\alpha\,v'+3\,\alpha\,u'_x\right)\psi_{x}^{'-}
\\&
\ \ \ \ \ \ \ \ \
+4\,\psi_{xxx}^{'-}+\left(6\,\alpha\,u'-\frac{12\,\beta}{\alpha}\right)\psi_{xx}^{'-},
\end{align}\label{Gar5.3}
\end{subequations}
and
\begin{subequations}
\begin{align}
&\psi_{y}^{'+}=\psi_{xx}^{'+}-\left(\alpha\,u'-\frac{2\,\beta}{\alpha}\right)\psi_{x}^{'+},
\\ \nonumber
&\psi_{t}^{'+}=
\left(\frac{12\,\beta^2}{\alpha^2}-6\,\beta\,u'+\frac{3}{2}\,\alpha^2\,u'^2-3\,\alpha\,v'-3\,\alpha\,u'_x\right)\psi_{x}^{'+}
\\& \ \ \ \ \ \ \ \ \
+4\,\psi_{xxx}^{'+}-\left(6\,\alpha\,u'-\frac{12\,\beta}{\alpha}\right)\psi_{xx}^{'+}.
\end{align}\label{Gar5.4}
\end{subequations}\hspace{-1.5mm}Substituting Eqns.~(\ref{Gar2.4})$-$(\ref{Gar2.5}) and
Ansatzs (\ref{Gar5.1})$-$(\ref{Gar5.2}) into
Eqns.~(\ref{Gar5.3})$-$(\ref{Gar5.4}) and equating to zero the
coefficients of like powers $\phi$ and $\varphi$, with symbolic
computation,  yield the following set of equations:
\begin{eqnarray}
&&f_{1x}=Bf_1-\frac{\phi_x\,\psi_1^-}{B},\label{Gar5.5}
\\&&f_{1y}=\frac{\psi_y^-}{\psi^-}\,f_1+\frac{\phi_x}{B}\left(-\psi_{1y}^--A\,\psi_{1x}^-\right),\label{Gar5.6}
\\&&f_{1t}=\frac{\psi_t^-}{\psi^-}\,f_1+\frac{\phi_x\,\psi_{1x}^-}{B}\left(-\frac{\psi_{1t}^-}{\psi_{1x}^-}-4\,A\,\frac{\psi_{1y}^-}{\psi_{1x}^-}
+4\,A\,\frac{\beta}{\alpha}-2\,\alpha\,A\,u-4\,A\,\tau_2-4\,AB\right),\label{Gar5.7}
\\&&f_2=0,\label{Gar5.8}\end{eqnarray} and
\begin{eqnarray}
&&g_{2x}=A\,g_2-\frac{\varphi_x\,\psi_1^+}{A},\label{Gar5.9}
\\&&g_{2y}=\frac{\psi_y^+}{\psi^+}\,g_2+\frac{\varphi_x}{A}\left(-\psi_{1y}^++B\,\psi_{1x}^+\right),\label{Gar5.10}
\\&&g_{2t}=\frac{\psi_t^+}{\psi^+}\,g_2+\frac{\varphi_x\,\psi_{1x}^+}{A}\left(-\frac{\psi_{1t}^+}{\psi_{1x}^+}+4\,B\,\frac{\psi_{1y}^+}{\psi_{1x}^+}
-4\,B\,\frac{\beta}{\alpha}+2\,\alpha\,B\,u+4\,B\,\tau_1-4\,AB\right)\hspace{-1.2mm}\label{Gar5.11},
\\&&g_1=0\label{Gar5.12}.
\end{eqnarray}
From Eqns.~(\ref{Gar5.5})$-$(\ref{Gar5.7}) and
(\ref{Gar5.9})$-$(\ref{Gar5.11}), $f_1$ and $g_2$ can be determined
as
\begin{eqnarray}
&&f_1=-\psi^-\Delta(\psi_1^-,\psi^+),\label{Gar5.13}
\\&&g_2=-\psi^+\Omega\,(\psi^-,\psi_1^+).\label{Gar5.14}
\end{eqnarray}
Now, the relation between the new and old eigenfunctions has been
constructed. Therefore, we arrive at the Darboux transformation for
System~(\ref{Gar2.1}) in the form
\begin{eqnarray}
&&u'=u+\frac{2}{\alpha}\left\{\rm{log}\left[\frac{\Delta(\psi^-,\psi^+)}{\Omega\,(\psi^-,\psi^+)}\right]\right\}_x,\label{Gar5.15}
\\&&v'=v+\frac{2}{\alpha}\left\{\rm{log}\left[\frac{\Delta(\psi^-,\psi^+)}{\Omega\,(\psi^-,\psi^+)}\right]\right\}_y,\label{Gar5.16}
\\&&\psi^{'-}=\psi_1^--\psi^-\,\frac{\Delta(\psi_1^-,\psi^+)}{\Delta(\psi^-,\psi^+)},\label{Gar5.17}
\\&&\psi^{'+}=\psi_1^+-\psi^+\,\frac{\Omega\,(\psi^-,\psi_1^+)}{\Omega\,(\psi^-,\psi^+)}.\label{Gar5.18}
\end{eqnarray}
\\
\noindent\textbf{7. Iteration of binary Darboux transformation and
Grammian solutions}
\\\hspace*{\parindent}In
this section,  we will perform the binary Darboux transformation $N$
times and progressively generate the analytical Grammian solutions.
After the second iteration of the binary Darboux transformation, the
new potential functions and eigenfunctions are expressed as
\renewcommand{\theequation}{7.\arabic{equation}}
\setcounter{equation}{0}
\begin{eqnarray}
&&u[2]=u'+\frac{2}{\alpha}\left(\frac{\phi_x'}{\phi'}-\frac{\varphi_x'}{\varphi'}\right)=u+\frac{2}{\alpha}
\left\{\rm{log}\left(\frac{\Phi[2]}{\Sigma[2]}\right)\right\}_x,\label{Gar6.1}
\\&&v[2]=v'+\frac{2}{\alpha}\left(\frac{\phi_y'}{\phi'}-\frac{\varphi_y'}{\varphi'}\right)=v+\frac{2}{\alpha}
\left\{\rm{log}\left(\frac{\Phi[2]}{\Sigma[2]}\right)\right\}_y,\label{Gar6.2}
\end{eqnarray}
with
\begin{eqnarray}
&&\Phi[2]=\phi\,\phi'=\Delta(\psi^-,\psi^+)\,\Delta(\psi^{'-},\psi^{'+}),\label{Gar6.3}
\\&&\Sigma[2]=\varphi\,\varphi'=\Omega\,(\psi^-,\psi^+)\,\Omega\,(\psi^{'-},\psi^{'+}),\label{Gar6.4}
\end{eqnarray}where $\phi'$ and $\varphi'$ are the new singular manifolds for
$(u',v')$, while ($\phi, \varphi$)  and ($\Phi[2],\Sigma[2]$) are
the first and second iterative $\tau$-functions in the Hirota
method~\cite{P93,RM95,PG94}. From Eqns.~(\ref{Gar6.3}) and
(\ref{Gar6.4}), one can immediately calculate that
\begin{eqnarray}
&&\Phi[2]=\begin{vmatrix}
\Delta(\psi_1^-,\psi_1^+) &\Delta(\psi^-_2,\psi_1^+)\\
\Delta(\psi_1^-,\psi_2^+)&\Delta(\psi^-_2,\psi^+_2)\end{vmatrix},\label{Gar6.5}
\\&&\Sigma[2]=\begin{vmatrix}
\Omega\,(\psi_1^-,\psi_1^+) &\Omega\,(\psi^-_2,\psi_1^+)\\
\Omega\,(\psi_1^-,\psi^+_2)&\Omega\,(\psi^-_2,\psi^+_2)\end{vmatrix}.\label{Gar6.6}
\end{eqnarray}
Taking $\psi_i^-$ and $\psi_i^+$ $(\,i=1,2,3\,)$ as the solutions of
the Lax pairs ~(\ref{Gar3.32}) and (\ref{Gar3.33}) for $(u,v)$ and
iterating the Darboux transformation, we can get the following
results:

\begin{eqnarray}
&&u[3]=u+\frac{2}{\alpha}
\left\{\rm{log}\left(\frac{\Phi[3]}{\Sigma[3]}\right)\right\}_x,\label{Gar6.7}
\\&&v[3]=v+\frac{2}{\alpha}
\left\{\rm{log}\left(\frac{\Phi[3]}{\Sigma[3]}\right)\right\}_y,\label{Gar6.8}
\end{eqnarray}
with
\begin{eqnarray}
&&\Phi[3]=\begin{vmatrix}
\Delta(\psi_1^-,\psi_1^+) &\Delta(\psi_1^-,\psi_2^+)&\Delta(\psi_1^-,\psi_3^+)\\
\Delta(\psi_2^-,\psi_1^+)&\Delta(\psi_2^-,\psi_2^+)&\Delta(\psi_2^-,\psi_3^+)\\
\Delta(\psi_3^-,\psi_1^+)&\Delta(\psi_3^-,\psi_2^+)&\Delta(\psi_3^-,\psi_3^+)\end{vmatrix},\label{Gar6.9}
\\&&\Sigma[3]=\begin{vmatrix}
\Omega\,(\psi_1^-,\psi_1^+) &\Omega\,(\psi_1^-,\psi_2^+)&\Omega\,(\psi_1^-,\psi_3^+)\\
\Omega\,(\psi_2^-,\psi_1^+)&\Omega\,(\psi_2^-,\psi_2^+)&\Omega\,(\psi_2^-,\psi_3^+)\\
\Omega\,(\psi_3^-,\psi_1^+)&\Omega\,(\psi_3^-,\psi_2^+)&\Omega\,(\psi_3^-,\psi_3^+)\end{vmatrix}.\label{Gar6.10}
\end{eqnarray}
Following the same procedure above, we iterate the Darboux
transformation $N$ times with symbolic computation and obtain
\begin{eqnarray}
&&u[N]=u+\frac{2}{\alpha}
\left\{\rm{log}\left(\frac{\Phi[N]}{\Sigma[N]}\right)\right\}_x,\label{Gar6.11}
\\&&v[N]=v+\frac{2}{\alpha}
\left\{\rm{log}\left(\frac{\Phi[N]}{\Sigma[N]}\right)\right\}_y,\label{Gar6.12}
\end{eqnarray}
with
\begin{eqnarray}
&&\Phi[N]=\begin{vmatrix}
\Delta(\psi_1^-,\psi_1^+) &\Delta(\psi_1^-,\psi_2^+)&\cdots&\Delta(\psi_1^-,\psi_N^+)\\
\Delta(\psi_2^-,\psi_1^+)&\Delta(\psi_2^-,\psi_2^+)&\cdots&\Delta(\psi_2^-,\psi_N^+)
\\ \vdots & \vdots &\ddots&\vdots
\\\Delta(\psi_N^-,\psi_1^+)&\Delta(\psi_N^-,\psi_2^+)&\cdots&\Delta(\psi_N^-,\psi_N^+)\end{vmatrix},\label{Gar6.13}
\\&&\Sigma[N]=\begin{vmatrix}
\Omega\,(\psi_1^-,\psi_1^+) &\Omega\,(\psi_1^-,\psi_2^+)&\cdots&\Omega\,(\psi_1^-,\psi_N^+)\\
\Omega\,(\psi_2^-,\psi_1^+)&\Omega\,(\psi_2^-,\psi_2^+)&\cdots&\Omega\,(\psi_2^-,\psi_N^+)
\\ \vdots & \vdots &\ddots&\vdots
\\ \Omega\,(\psi_N^-,\psi_1^+)&\Omega\,(\psi_N^-,\psi_2^+)&\cdots&\Omega\,(\psi_N^-,\psi_N^+)\end{vmatrix},\label{Gar6.14}
\end{eqnarray}
where $\psi_i^-$ and $\psi_i^+$ $(i=1,2,\cdots,N)$ satisfy Lax pairs
~(\ref{Gar3.32}) and (\ref{Gar3.33}).

In illustration, we take $u=v=0$ as the seed solutions for Lax pairs
~(\ref{Gar3.32}) and (\ref{Gar3.33}), yielding
\begin{eqnarray}
&&\psi_1^-={\rm{exp}}\left\{k_1\,x+l_1\,y+w_1\,t\right\},
\\&&\psi_1^+={\rm{exp}}\left\{p_1\,x+m_1\,y+n_1\,t\right\},
\\&&\psi_2^-={\rm{exp}}\left\{k_2\,x+l_2\,y+w_2\,t\right\},
\\&&\psi_2^+={\rm{exp}}\left\{p_2\,x+m_2\,y+n_2\,t\right\},
\end{eqnarray}
where $l_i=\frac{2\beta}{\alpha}k_i-k_i^2$,
$w_i=4k_i^3-\frac{12\beta}{\alpha}k_i^2+\frac{12\beta^2}{\alpha^2}k_i$,
$m_i=\frac{2\beta}{\alpha}p_i+p_i^2$,
$n_i=4p_i^3+\frac{12\beta}{\alpha}p_i^2+\frac{12\beta^2}{\alpha^2}p_i$
with $k_i$ and $p_i$ $(\,i=1,2\,)$ as arbitrary constants.

Integration of Eqns.~(\ref{Gar4.15})$-$(\ref{Gar4.20}) with respect
to $x$, $y$ and $t$ results in
\begin{eqnarray}
\Delta(\psi_1^-,\psi_1^+)=\frac{k_1}{k_1+p_1}\psi_1^-\,\psi_1^++\delta_1,
\end{eqnarray}
\begin{eqnarray}
\\&&\Delta(\psi_2^-,\psi_2^+)=\frac{k_2}{k_2+p_2}\psi_2^-\,\psi_2^++\delta_2,
\\&&\Delta(\psi_1^-,\psi_2^+)=\frac{k_1}{k_1+p_2}\psi_1^-\,\psi_2^++\delta_3,
\\&&\Delta(\psi_2^-,\psi_1^+)=\frac{k_2}{k_2+p_1}\psi_2^-\,\psi_1^++\delta_4,
\\&&\Omega\,(\psi_1^-,\psi_1^+)=\frac{p_1}{k_1+p_1}\psi_1^-\,\psi_1^+-\delta_1,
\\&&\Omega\,(\psi_2^-,\psi_2^+)=\frac{p_2}{k_2+p_2}\psi_2^-\,\psi_2^+-\delta_2,
\\&&\Omega\,(\psi_1^-,\psi_2^+)=\frac{p_1}{k_1+p_2}\psi_1^-\,\psi_2^+-\delta_3,
\\&&\Omega\,(\psi_2^-,\psi_1^+)=\frac{p_2}{k_2+p_1}\psi_2^-\,\psi_1^+-\delta_4,
\end{eqnarray}
where $\delta_i$ $(\,i=1,2,3,4\,)$ are all arbitrary constants.
Particularly, setting $\delta_1=\delta_2=1$ and
$\delta_3=\delta_4=0$, the $\tau$-functions can be expressed as
follows:
\begin{eqnarray}
&&\hspace{-4.5mm}\phi=1+k_1\,F_1,\label{Gar6.27}
\\&&\hspace{-4.5mm}\varphi=p_1\,F_1-1,\label{Gar6.28}
\\&&\hspace{-0.5cm}\Phi[2]=1+k_1\,F_1+k_2\,F_2+k_1k_2\,R\,F_1\,F_2,\label{Gar6.29}
\\&&\hspace{-0.5cm}\Sigma[2]=1-p_1\,F_1-p_2\,F_2+p_1p_2\,R\,F_1\,F_2,\label{Gar6.30}
\end{eqnarray}
where
\begin{eqnarray}
&&F_1={\rm{exp}}\left\{(k_1+p_1)\,x+(l_1+m_1)\,y+(w_1+n_1)\,t+\theta_1\right\},
\\&&F_2={\rm{exp}}\left\{(k_2+p_2)\,x+(l_2+m_2)\,y+(w_2+n_2)\,t+\theta_2\right\},
\\&&\hspace{1mm}R=\frac{(k_1-k_2)(p_1-p_2)}{(k_1+p_1)(k_2+p_2)},
\\&&e^{-(k_1+p_1)\theta_1}=k_1+p_1,
\\&&e^{-(k_2+p_2)\theta_2}=k_2+p_2.
\end{eqnarray}
It is of interest to note that
Expressions~(\ref{Gar6.27})$-$(\ref{Gar6.30}) are very similar to
the forms of soliton solutions obtained with  the Hirota
method~\cite{MKdV87}. Substituting
Expressions~(\ref{Gar6.27})$-$(\ref{Gar6.30}) into
Eqns.~(\ref{Gar5.15}), (\ref{Gar5.16}), (\ref{Gar6.1}) and
(\ref{Gar6.2}), one can get the one and two soliton solutions in the
sense of Refs.~\cite{PP97,MKdV87,Den06,Den07}.
\\
\\
\noindent\textbf{8. Conclusions}
\\\hspace*{\parindent}The singular manifold method from the Painlev\'{e} analysis plays a vital
role in investigating many important integrable properties for the
NPDEs. In this paper, we have successfully applied the
two-singular-manifold method to the (2+1)-dimensional Gardner
equation and derived its Hirota bilinear form, bilinear B\"{a}cklund
transformation, Lax pairs, as well as binary Darboux transformation.
With the help of symbolic computation, we have performed the
$N$-time iterative algorithm of binary Darboux transformation to
generate the $N \times N$ Grammian solution.
\\
\\{\bf Acknowledgments}

We express our thanks to Prof.\ Y.\ T.\ Gao and Ms.\ X.\ H.\ Meng
for their valuable comments. This work has been supported by the Key
Project of Chinese Ministry of Education (No.\ 106033), by the
Specialized Research Fund for the Doctoral Program of Higher
Education (No.\ 20060006024), Chinese Ministry of Education, and by
the National Natural Science Foundation of China under Grant No.\
60372095.
\\
\\
\noindent\textbf{Appendix: Identity properties of the bilinear
operator}
\\\hspace*{\parindent}The following identities are used in the
derivation of the bilinear B\"{a}cklund transformation.
\renewcommand{\theequation}{A.\arabic{equation}}
\setcounter{equation}{0}
\begin{eqnarray}
& &\hspace{-15mm}D_x\,c_0\,a(x)\cdot a(x)=D_x\,a(x)\cdot
c_0\,a(x)=0, \hspace{1cm}(\,c_0=const.\,),\label{GarA1}
\\& &\hspace{-15mm}(D_x\,a\cdot b)\,c\,d-a\,b\,(D_x\,c\cdot d)=
(D_x\,a\cdot c)\,b\,d-a\,c(D_x b\cdot
d)=D_x(a\,d)\cdot(b\,c),\label{GarA2}
\\& &\hspace{-15mm}(D_x^2\,a\cdot b)\,c\,d-a\,b\,(D_x^2\,c\cdot d)=
(D_x^2\,a\cdot c)\,b\,d-a\,c(D_x^2\,b\cdot
d)-2\,D_x\,a\,d\cdot(D_x\,b\cdot c),\label{GarA3}
\\& &\hspace{-15mm}(D_x^3\,a\cdot b)\,c\,d-a\,b\,(D_x^3\,c\cdot d)=
(D_x^3\,a\cdot c)\,b\,d-a\,c(D_x^3\,b\cdot d)-3\,D_x(D_x\,a\cdot
d)\cdot (D_x\,b\cdot c),\label{GarA4}
\\ & &\hspace{-15mm}(D_xD_y\,a\cdot b)\,c\,d-a\,b(\,D_xD_y\,c\cdot
d)=(D_xD_y\,a\cdot c)\,b\,d-a\,c\,(\,D_xD_y\,b\cdot d)\nonumber
\\& &
\hspace{5cm}-D_x\,(a\,d)\cdot(D_y\,b\cdot c)-D_y(a\,d)\cdot (D_x\,b
\cdot c)\label{GarA5}.
\end{eqnarray}

\end{document}